\documentclass[12pt]{article}
\usepackage{amsmath,amsfonts,graphicx,yfonts,mathabx}
\usepackage[brazil]{babel}

\newcommand{\be}{\begin{equation}}
\newcommand{\ee}{\end{equation}}
\newcommand{\bea}{\begin{eqnarray}}
\newcommand{\eea}{\end{eqnarray}}
\newcommand{\mt}[1]{\textrm{\tiny #1}}
\renewcommand{\title}[1]{\vbox{\center\LARGE{#1}}\vspace{3mm}}
\renewcommand{\author}[1]{\vbox{\center#1}\vspace{3mm}}
\newcommand{\address}[1]{\vbox{\center\em#1}}
\newcommand{\email}[1]{\vbox{\center\tt#1}\vspace{3mm}}

\begin{document}
\begin{titlepage}
\begin{center}
\rightline{\tt}
\vskip 2.5cm
\title{Grandezas f\'isicas e an\'alise dimensional:\\
da mec\^anica \`a gravidade qu\^antica}
\vskip .6cm
\author{Diego Trancanelli}
\vskip -.5cm 
\address{Instituto de F\'isica, Universidade de S\~ao Paulo\\ 
05314-970 S\~ao Paulo, Brasil}
\vskip -.1cm 
\email{dtrancan@usp.br}
\end{center}
\vskip 3cm

\abstract{\noindent  
Entre os primeiros conceitos que alunos de gradua\c c\~ao em f\'isica encontram nos seus estudos, h\'a os de grandezas e dimens\~oes f\'isicas. Neste artigo pedag\'ogico de revis\~ao, usando a poderosa ferramenta da an\'alise dimensional, vou partir desses conceitos para uma viagem atrav\'es de v\'arios ramos da f\'isica te\'orica, da mec\^anica at\'e a gravidade qu\^antica. Entre outras coisas, vou discutir um pouco sobre as constantes fundamentais da Natureza, o chamado ``cubo da f\'isica'' e o sistema de unidades naturais.\\
{\bf Palavras-chave:} an\'alise dimensional, constantes fundamentais, unidades naturais.}
\vfill
\end{titlepage}

\section{Introdu\c c\~ao}

O conceito de {\it grandeza f\'isica} \'e um dos mais b\'asicos entre os conceitos encontrados pelos alunos que come\c cam o estudo da f\'isica, sendo, talvez, at\'e axiom\'atico. Grandezas f\'isicas s\~ao usadas para descrever fen\^omenos ou propriedades de sistemas e s\~ao  caracterizadas por terem {\it dimens\~oes} como, por exemplo, dimens\~oes de massa, for\c ca ou energia.

As dimens\~oes podem ser {\it primitivas} ou {\it derivadas}.  As dimens\~oes primitivas s\~ao tr\^es: a {\bf massa} (que indicamos com a letra $M$), o {\bf comprimento} (que indicamos com a letra $L$, da palavra ingl\^es {\it length}) e o {\bf tempo} (que indicamos com a letra $T$). Todas as outras dimens\~oes s\~ao derivadas, ou seja, podem ser expressas em termos das tr\^es primitivas. O leitor \'e convidado a abrir um livro-texto b\'asico qualquer, como \cite{moyses}, e verificar isso explicitamente.%
\footnote{Contudo, \`as vezes \'e \'util considerar outras dimens\~oes como se fossem primitivas, para motivos pr\'aticos. Por exemplo, poder\'iamos considerar a temperatura e a carga el\'etrica tamb\'em como primitivas, mas \'e importante ressaltar que isso n\~ao \'e estritamente necess\'ario. De fato, a temperatura tem as mesmas dimens\~oes de energia e a carga el\'etrica tem dimens\~oes de $\sqrt{M L^3T^{-2}}$. Em outras palavras, a constante de Boltzmann $k_B$ e a constante de Coulomb $1/4\pi\varepsilon_0$ s\~ao simplesmente fatores de convers\~ao e n\~ao exercem um papel fundamental.}
Por exemplo, uma acelera\c c\~ao tem dimens\~oes de comprimento sobre tempo ao quadrado, $L T^{-2}$, enquanto a energia tem dimens\~oes de massa vezes comprimento ao quadrado sobre tempo ao quadrado, $M L^2T^{-2}$. 

No que segue, indicamos as dimens\~oes de uma grandeza f\'isica $X$ com colchetes: $[X]$. As dimens\~oes de $X$ s\~ao, ent\~ao, dadas por uma certa combina\c c\~ao das dimens\~oes primitivas
\be
[X]=M^\alpha L^\beta T^\gamma\,,
\label{dimX}
\ee
onde $\alpha$, $\beta$ e $\gamma$ s\~ao n\'umeros inteiros ou racionais, positivos ou negativos.\footnote{A depend\^encia de $[X]$ em $M$, $L$ e $T$ n\~ao pode ser mais complicada que o mon\^omio em (\ref{dimX}) por raz\~oes que logo ficar\~ao claras.} Nos exemplos anteriores, sobre acelera\c c\~ao e energia, esses n\'umeros s\~ao, respectivamente, $(\alpha=0,\beta=1,\gamma=-2)$ e  $(\alpha=1,\beta=2,\gamma=-2)$.  Al\'em das grandezas f\'isicas dimensionais, h\'a tamb\'em os n\'umeros puros, como $0$, $1$ ou $\pi$, que s\~ao {\it adimensionais}, ou seja, t\^em expoentes $(\alpha=0,\beta=0,\gamma=0)$.

\`As vezes, \'e \'util mudar de base de dimens\~oes. A nova base precisa ser composta de dimens\~oes independentes e precisa ser completa. Por exemplo, for\c ca ($F$), comprimento e tempo representam uma base alternativa leg\'itima, pois $F=MLT^{-2}$ cont\'em $M$. Por outro lado, velocidade ($V$), comprimento e tempo n\~ao seria uma base leg\'itima, pois $V=LT^{-1}$ n\~ao \'e independente das outras duas e n\~ao \'e poss\'ivel descrever massas nessa base. Veremos na Se\c c\~ao \ref{planck} que existe uma base muito conveniente de dimens\~oes, chamada justamente de base ``natural''.

Depois dessa breve introdu\c c\~ao, estamos prontos para discutir o que \'e a an\'alise dimensional. 


\section{An\'alise dimensional}
\label{proc}

A an\'alise dimensional \'e a {\it arte} de antecipar como uma certa grandeza vai depender de outras grandezas presentes em um problema, sem fazer a conta detalhada, ou seja, sem ter que resolver explicitamente as equa\c c\~oes que definem o problema. A an\'alise dimensional \'e uma ferramenta extremamente poderosa na caixa de ferramentas de um f\'isico, te\'orico ou experimental, e \'e muito \'util para v\'arias tarefas, entre elas:
\begin{itemize}
\item
{\bf Simplificar problemas.} A an\'alise dimensional permite entender quais s\~ao as combina\c c\~oes de grandezas f\'isicas relevantes para o problema abordado. Abaixo, vamos demonstrar esse ponto atrav\'es de v\'arios exemplos concretos.

\item
{\bf Desenvolver intui\c c\~ao f\'isica.} Uma das habilidades mais importantes para um f\'isico \'e ter {\it intui\c c\~ao}, ou seja, a capacidade de imaginar o comportamento de um sistema ao mudarmos os par\^ametros que o controlam, sem fazer contas detalhadas. A an\'alise dimensional \'e  uma das t\'ecnicas mais \'uteis para alcan\c car isso. 

\item
{\bf Detectar erros.} Talvez essa seja a aplica\c c\~ao mais imediatamente relevante para os alunos. Uma boa parte dos erros na solu\c c\~ao de problemas pode ser detectada simplesmente checando se as dimens\~oes do resultado final s\~ao apropriadas para a grandeza que se pretende calcular.
\end{itemize}

O princ\'ipio b\'asico da an\'alise dimensional \'e o de {\bf homogeneidade dimensional}  de qualquer equa\c c\~ao. Isso significa que todos os termos da equa\c c\~ao t\^em que  ter as mesmas dimens\~oes. Enquanto podemos combinar (multiplicando ou dividindo) grandezas f\'isicas com dimens\~oes diferentes, somente podemos comparar (somando ou subtraindo) grandezas f\'isicas com as mesmas dimens\~oes. Outra maneira de falar isso, em termos mais populares, \'e que s\'o podemos comparar bananas com bananas e laranjas com laranjas, e n\~ao bananas com laranjas.

O procedimento para entender como uma grandeza $Y$ depende das outras grandezas do problema usando a an\'alise dimensional \'e o seguinte:
\begin{enumerate}
\item
Come\c camos escrevendo as dimens\~oes $[Y]$ de $Y$ em termos das tr\^es dimens\~oes primitivas
\be
[Y]=M^\alpha L^\beta T^\gamma\,.
\label{dimY}
\ee
Isso determina os expoentes $(\alpha,\beta,\gamma)$.

\item
Tentamos identificar as grandezas $X_i$ ($i=1,2,\ldots$) das quais $Y$ pode depender. Essa \'e a parte mais complicada do procedimento e \'e gra\c cas a ela que, como escrevi acima, a an\'alise dimensional pode ser considerada uma arte! Por exemplo, ser\'a que o per\'iodo de um p\^endulo simples depende da temperatura da sala onde ele se encontra? Veremos v\'arios exemplos nas pr\'oximas se\c c\~oes como decidir isso na pr\'atica.

\item
Escrevemos $Y$ como um mon\^omio destas grandezas $X_i$, com uma quantidade adimensional $C$ na frente
\be
Y=C X_1^{a_1} X_2^{a_2}\ldots\,,
\label{Y}
\ee
onde $a_1,a_2,\ldots$ s\~ao expoentes a serem determinados.

\item
Comparamos a equa\c c\~ao acima com (\ref{dimY}) e resolvemos para os expoentes $a_1, a_2,\ldots$
\be
[X_1^{a_1}][X^{a_2}_2]\ldots=M^\alpha L^\beta T^\gamma\,,
\ee
lembrando que $[C]=1$, pois $C$ \'e adimensional. Por exemplo, consideramos o caso em que haja somente tr\^es grandezas relevantes para o problema: $X_1$, $X_2$ e $X_3$. Imaginemos tamb\'em que $X_1$ tenha dimens\~oes dadas pela escolha de expoentes $(\alpha_1,\beta_1,\gamma_1)$, $X_2$ tenha $(\alpha_2,\beta_2,\gamma_2)$ e, finalmente, $X_3$ tenha $(\alpha_3,\beta_3,\gamma_3)$. O sistema de equa\c c\~oes a ser resolvido resulta ser
\bea
& a_1 \alpha_1+a_2\alpha_2+a_3 \alpha_3=\alpha\,, \cr
& a_1 \beta_1+a_2\beta_2+a_3\beta_3=\beta\,, \cr
& a_1 \gamma_1+a_2\gamma_2+a_3 \gamma_3=\gamma\,.
\eea
Encontrando as solu\c c\~oes para $a_1$, $a_2$ e $a_3$ podemos, portanto, determinar a depend\^encia de $Y$ em rela\c c\~ao a $X_1$, $X_2$ e $X_3$.

\end{enumerate}

Seguem agora tr\^es observa\c c\~oes importantes. A primeira \'e sobre os argumentos de fun\c c\~oes, como, por exemplo, fun\c c\~oes trigonom\'etricas ou a fun\c c\~ao exponencial. Esses argumentos t\^em que ser adimensionais. \'E f\'acil se convencer disso observando que na expans\~ao destas fun\c c\~oes, devido ao princ\'ipio da homogeneidade dimensional, todos os termos t\^em que ter as mesmas dimens\~oes. Por exemplo,
\be
\cos x =1-\frac{x^2}{2} +\frac{x^4}{24}+\ldots\,,
\ee
logo $x$ deve ser adimensional, sendo o primeiro termo do lado direito da equa\c c\~ao, 1, adimensional. 

A segunda observa\c c\~ao se chama {\it Teorema-$\Pi$ de Buckingham} \cite{Pi} e \'e uma contagem muito simples de quantas combina\c c\~oes adimensionais independentes podem ser formadas a partir das grandezas do problema. Se houver $n$ grandezas, o n\'umero $r$ de combina\c c\~oes adimensionais que pode ser obtido \'e dado por\footnote{O 3 nessa equa\c c\~ao se deve ao fato de ter sido escolhido trabalhar com tr\^es dimens\~oes primitivas. Incluindo, por exemplo, temperatura e carga el\'etrica como dimens\~oes primitivas, al\'em de $M$, $L$ e $T$, devemos substituir o 3 por um 5.}
\be
r=n-3\,.
\ee 
Chamaremos essas combina\c c\~oes adimensionais de $\Pi_a$, com $a=1,\ldots, r$. \'E f\'acil ver (e veremos, logo, em uma s\'erie de exemplos) que, para $r\le 1$, a quantidade $C$  introduzida antes em (\ref{Y}) \'e uma constante (como, por exemplo, $2\pi$ ou $1/2$), mas, para $r\ge 2$, a $C$ ser\'a uma fun\c c\~ao de algumas das combina\c c\~oes $\Pi_a$. 

Finalmente, a \'ultima observa\c c\~ao \'e que, quando tomamos limites, devemos tomar cuidado para que o limite fa\c ca sentido, ou seja, para estarmos comparando, no limite, quantidades com as mesma dimens\~oes. Em particular, falar que $x\ll 1$ ou $x\gg 1$, somente faz sentido se $x$ for um n\'umero puro como 1. Voltaremos a esse ponto mais adiante.

Vamos testar logo essas no\c c\~oes em alguns exemplos simples de mec\^anica cl\'assica.


\subsubsection*{Exemplo 1: Per\'iodo de um p\^endulo}

Podemos entender como o per\'iodo de um p\^endulo depende dos outros par\^ametros do problema usando a an\'alise dimensional. Primeiramente, notamos que h\'a cinco grandezas relevantes no problema:\footnote{Como mencionado antes, essa \'e a parte dif\'icil da an\'alise: entender o que \'e relevante e o que n\~ao \'e relevante (como, por exemplo, a temperatura do ambiente em volta do p\^endulo).} o per\'iodo $\tau$ (com dimens\~oes de $T$), o comprimento da corda $\ell$ (com dimens\~oes de $L$), a massa suspensa $m$ (com dimens\~oes de $M$), a acelera\c c\~ao da gravidade $g$ (com dimens\~oes de $LT^{-2}$) e o \^angulo $\theta_0$ do deslocamento inicial do p\^endulo (que \'e adimensional).\footnote{\^Angulos precisam ser adimensionais devido \`a periodicidade: $\theta_0$ deve ser identificado com $\theta_0+2\pi$, e portanto $\theta_0$ deve ter as mesmas dimens\~oes de $2\pi$.}
\begin{figure}
\begin{center}
\setlength{\unitlength}{1cm}
\includegraphics[scale=.4]{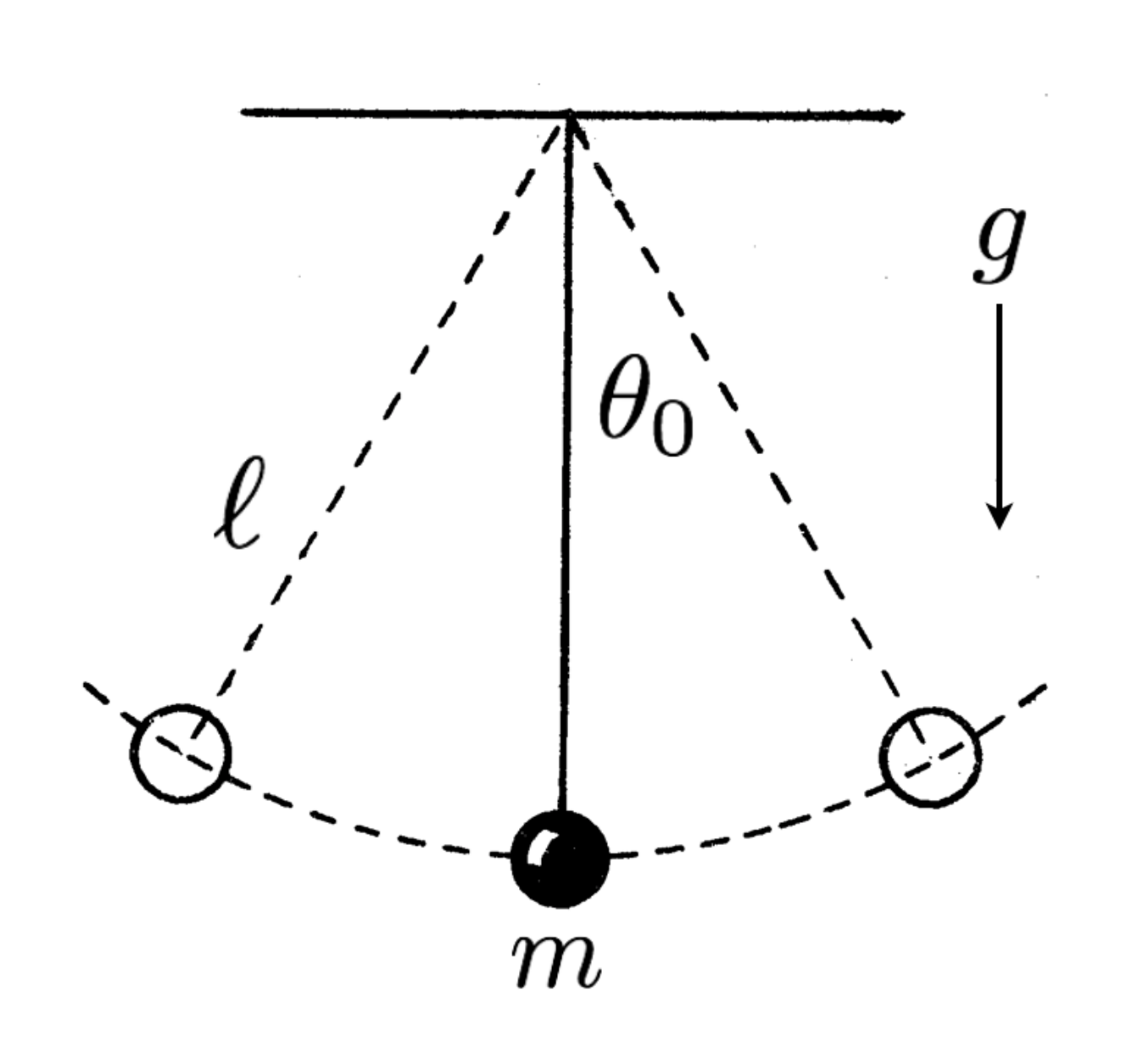} 
\end{center}
\caption{\small
P\^endulo simples.}
\label{pendulo}
\end{figure}
Ver a Figura \ref{pendulo}. Conforme o teorema-$\Pi$, haver\'a $r=5-3=2$ combina\c c\~oes adimensionais. A primeira combina\c c\~ao \'e, claramente, dada por $\Pi_1=\theta_0$, enquanto a segunda pode ser obtida combinando $\tau$, $\ell$, $m$ e $g$.  Usando o procedimento descrito acima, podemos escrever
 \bea
 \tau = C(\theta_0) \ell^{a_1} m^{a_2} g^{a_3}\,,
 \eea
 e, em termos das dimens\~oes,
 \bea
 [\tau]= [C(\theta_0)] [\ell]^{a_1} [m]^{a_2} [g]^{a_3}\quad \to \quad T=L^{a_1+a_3}M^{a_2}T^{-2a_3}\,, 
 \eea
que tem solu\c c\~ao \'unica dada por $a_1=\frac{1}{2}$, $a_2=0$ e $a_3=-\frac{1}{2}$. Logo
\be
\tau=C(\theta_0)\sqrt{\frac{\ell}{g}}\,.
\ee
Essa conta simples revela que o per\'iodo n\~ao pode depender da massa do p\^endulo! \'E claro que a an\'alise dimensional n\~ao \'e suficiente para determinar a fun\c c\~ao $C(\theta_0)$, que \'e a parte dif\'icil do problema e que pode ser obtida somente resolvendo a equa\c c\~ao diferencial do p\^endulo. Se estivermos interessados na compara\c c\~ao entre dois p\^endulos diferentes (de per\'iodos $\tau_1$ e $\tau_2$ e comprimentos $\ell_1$ e $\ell_2$), mas com o mesmo \^angulo inicial, essa fun\c c\~ao \'e de qualquer forma irrelevante, pois
\be
\frac{\tau_1}{\tau_2}=\sqrt{\frac{\ell_1}{\ell_2}}\,.
\ee
O leitor \'e convidado a repetir a mesma an\'alise para um oscilador harm\^onico de massa $m$ e constante el\'astica $k$.


\subsubsection*{Exemplo 2: Energia de uma corda vibrante}

Agora, queremos estimar a energia de uma corda de viol\~ao em vibra\c c\~ao. As vari\'aveis do problema s\~ao: a energia $E$, o comprimento da corda $\ell$, a amplitude das oscila\c c\~oes $A$ (parece claro que, quanto mais ampla \'e a oscila\c c\~ao, maior \'e a energia), a densidade linear de massa $\rho$ da corda (com dimens\~oes de $ML^{-1}$) e a tens\~ao  $s$ da corda (com dimens\~oes de for\c ca, $MLT^{-2}$). De novo h\'a  $r=5-3=2$ combina\c c\~oes adimensionais independentes que s\~ao $\Pi_1=E/As$ e $\Pi_2=\ell/A$. Portanto, a energia ser\'a
\be
E=As\, f(\ell/A)\,,
\ee
onde $f(\ell/A)$ \'e uma fun\c c\~ao (adimensional) que, claramente, n\~ao pode ser determinada usando a an\'alise dimensional. De qualquer forma, a an\'alise foi \'util para simplificar o problema, determinando a depend\^encia da energia com a tens\~ao da corda.  


\subsubsection*{Exemplo 3: Pot\^encia gasta por um mexedor de caf\'e}

Agora consideremos uma situa\c c\~ao do cotidiano: quanta energia gastamos para mexer o a\c c\'ucar no nosso caf\'e? \'E claro que esse problema \'e extremamente complicado e depende de muitos par\^ametros. A an\'alise dimensional, contudo, permite simplificar as coisas um pouco. Podemos imaginar que as grandezas (dimensionais) mais relevantes para o problema sejam: a pot\^encia $P$ gasta (ou seja, a energia gasta por unidade de tempo), a densidade volum\'etrica $\rho$ do caf\'e e a sua viscosidade $\mu$, o tamanho do mexedor (que podemos imaginar cil\'indrico, com di\^ametro $d$) e, enfim, a velocidade $v$ do movimento. Os leitores s\~ao convidados a escrever as dimens\~oes de todas essas grandezas. De novo, temos $r=5-3=2$ combina\c c\~oes $\Pi_a$ adimensionais. A primeira combina\c c\~ao depende das caracter\'isticas do fluido e se chama {\it n\'umero de Reynolds} $\mbox{Re}$\footnote{Esses n\'umeros s\~ao muito importantes no estudo da din\^amica dos fluidos, em particular no estudo da turbul\^encia, que acontece quando o fluido tem n\'umero de Reynolds grande.}
\be
\Pi_1\equiv \mbox{Re}=\frac{\rho v d}{\mu}
\,,
\ee
enquanto a segunda combina\c c\~ao se chama {\it n\'umero de pot\^encia} $N_\mt{p}$ e \'e dada por
\be
\Pi_2\equiv N_\mt{p}=\frac{P}{\rho v^3 d^5}
\,.
\ee
Vemos, de novo, que a an\'alise dimensional n\~ao foi suficiente para resolver o problema, como era esperado dada a complexidade do mesmo, mas, pelo menos, separou duas combina\c c\~oes adimensionais que ser\~ao relevantes. A pot\^encia gasta vai, de fato, depender das outras grandezas como segue
\be
P= \rho v^3 d^5 f(\mbox{Re})\,,
\ee
para uma certa fun\c c\~ao (adimensional) $f$ a ser determinada.


\subsubsection*{Exemplo 4: Energia liberada por bombas at\^omicas}

\begin{figure}
\begin{center}
\setlength{\unitlength}{1cm}
\includegraphics[scale=.13]{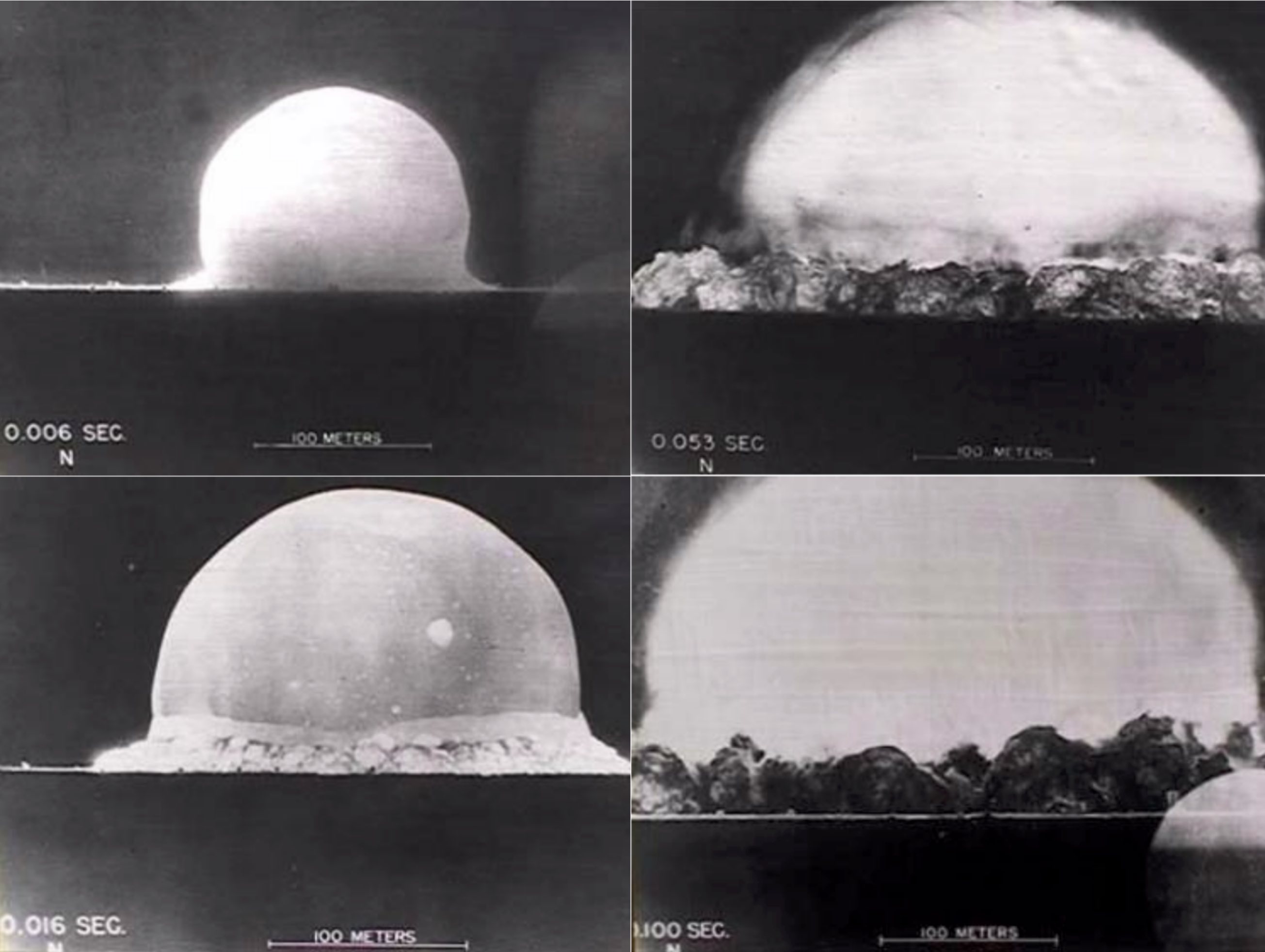} 
\end{center}
\caption{\small
Frentes de choque em explos\~oes de bombas at\^omicas.  Figura de \cite{bomb-ref}.}
\label{bomba}
\end{figure}

Uma outra aplica\c c\~ao interessante da an\'alise dimensional \'e estimar quanta energia \'e liberada na explos\~ao de uma bomba at\^omica \cite{bomb-ref}. Na d\'ecada de '50, o f\'isico ingl\^es G. I. Taylor estudou fotos de explos\~oes como a da Figura~\ref{bomba}.
Nessas fotos, podemos ler o tamanho da frente de choque da onda da explos\~ao (um comprimento que chamamos de $R$) em fun\c c\~ao do tempo $t$. A observa\c c\~ao importante \'e que podemos desprezar a press\~ao do ar em volta da explos\~ao, enquanto somente a densidade volum\'etrica $\rho$ do ar (com dimens\~oes de $ML^{-3}$) \'e importante. A energia liberada na explos\~ao deve, portanto, ser dada por
\be
E=C\,\frac{\rho R^5}{t^2}\,,
\ee
sendo $C$ uma constante adimensional. Extraindo $R(t)$ das fotos \'e poss\'ivel, portanto, conhecer $E$, a menos da constante $C$, que deve ser obtida com outros m\'etodos.
 
 
\subsubsection*{Exemplo 5: Velocidade de barcos de remos}

Finalmente, vamos tentar estimar como a velocidade de barcos de remos depende do n\'umero de remadores \cite{rowers0}  (ver tamb\'em \cite{rowers}). As grandezas relevantes s\~ao: o n\'umero $N$ de remadores, que \'e, claramente, adimensional, a \'area do barco submersa em baixo da \'agua $A$ (com dimens\~oes de $L^2$) e a velocidade $v$ do barco, sendo a densidade da \'agua irrelevante. A for\c ca de arrasto experimentada pelo barco \'e proporcional \`a velocidade ao quadrado\footnote{O s\'imbolo $\sim$ significa que n\~ao estamos interessados na depend\^encia precisa em todas as grandeza do problema (e nem em constantes n\'umericas adimensionais), mas somente em algumas grandezas selecionadas, como $v$ e $A$. Em particular, uma rela\c c\~ao indicada com $\sim$ pode at\'e ser inconsistente do ponto de vista dimensional, como, por exemplo, mais em baixo quando escrevemos $P\sim N$: aqui estamos suprimindo a pot\^encia individual de cada remador, que deixaria a rela\c c\~ao consistente. Em um certo sentido, este exemplo tem mais a ver com obter rela\c c\~oes de ``scaling'' que com a an\'alise dimensional.}
\be
F_\mt{arrasto}\sim v^2 A\,,
\ee
e a pot\^encia necess\'aria para compensar a perda de energia  \'e dada por
\be
P=F_\mt{arrasto}v \sim v^3 A\,.
\ee
O volume submerso pode ser estimado como sendo linear no n\'umero de remadores, $V\sim N$, implicando que $A\sim N^{2/3}$. Assumimos tamb\'em que todos os remadores remem com a mesma pot\^encia, $P\sim N$. Juntando todas essas f\'ormulas,
vemos que 
\be
v\sim N^{1/9}\,,
\ee
e, consequentemente, o tempo para cobrir uma dist\^ancia fixa vai como
\be
t\sim N^{-1/9}\,.
\ee
\'E interessante observar que essa predi\c c\~ao da an\'alise dimensional foi verificada \cite{rowers} usando os tempos dos recordes ol\'impicos nas especialidades de remos e a predi\c c\~ao \'e consistente com esses tempos com um erro de $\pm 1.5$ segundos!


\section{Unidades de medida}
\label{units}

O leitor atento deve ter reparado que, at\'e esse momento, tomei muito cuidado em nunca usar a express\~ao {\it unidades de medida}, somente usei o termo ``dimens\~oes''. A raz\~ao \'e que as duas coisas s\~ao muito diferentes: as dimens\~oes s\~ao propriedades intr\'insecas das grandezas f\'isicas, enquanto as unidades de medida s\~ao {\it conven\c c\~oes} usadas para descrever dimens\~oes (da mesma forma que um objeto \'e uma coisa diferente da palavra usada para descrev\^e-lo). Por exemplo, o {\it metro}  \'e uma unidade para medir um comprimento $L$, n\~ao o comprimento em si.

A introdu\c c\~ao de unidades de medida \'e necess\'aria para efetuar medidas, que s\~ao obtidas comparando uma certa grandeza f\'isica com uma grandeza de refer\^encia cujas dimens\~oes s\~ao as mesmas, especificando qual a grandeza de refer\^encia, como segue:
\be
\mbox{medida}=\frac{\mbox{grandeza f\'isica}}{\mbox{grandeza de refer\^encia}}\;\mbox{unidade}\,.
\ee
N\~ao irei discutir aqui os v\'arios sistemas de unidades de medida e nem a {\it metrologia}, que s\~ao assunto padr\~ao nos livros-texto, como por exemplo \cite{moyses}. 

\begin{figure}
\begin{center}
\setlength{\unitlength}{1cm}
\includegraphics[scale=.16]{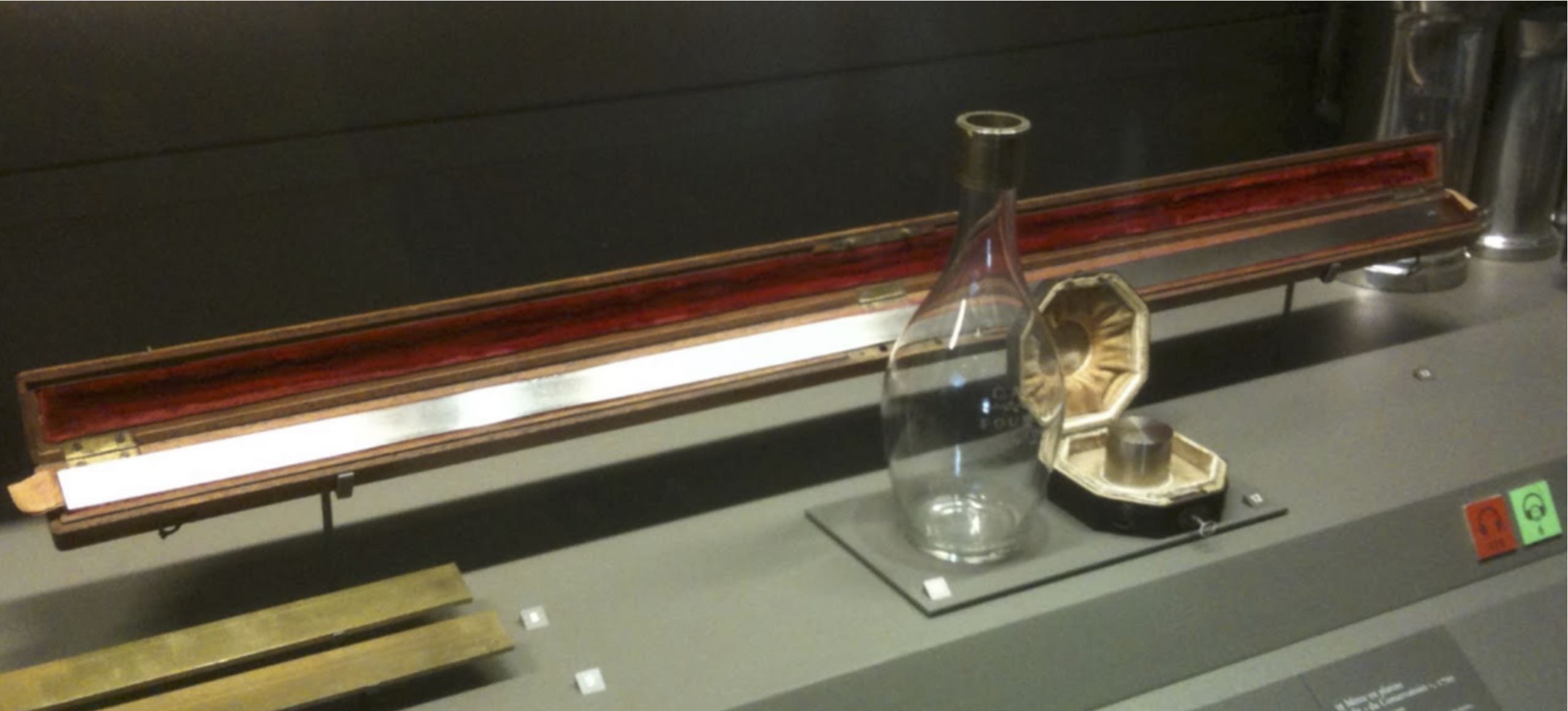} 
\end{center}
\caption{\small
Unidades de medida para massa, comprimento e volume no {\it Escrit\'orio Internacional de Pesos e Medidas} de S\`evres, na Fran\c ca.  Figura da Internet.}
\label{sevres}
\end{figure}

O ponto principal da discuss\~ao que segue \'e, ao inv\'es, ressaltar como esses sistemas de unidades s\~ao convencionais, dependendo de defini\c c\~oes derivadas da nossa experi\^encia local do universo. Por exemplo, o metro foi definido originalmente como uma certa fra\c c\~ao do comprimento dos meridianos terrestres e, embora essa defini\c c\~ao tenha sido refinada posteriormente, fica claro que \'e de natureza arbitr\'aria.  Na Figura~\ref{sevres}, s\~ao representadas as unidades de massa, comprimento e volume que, atualmente, encontram-se no museu de S\`evres.

\'E claro que, embora seja \'util introduzir essas conven\c c\~oes para fins pr\'aticos, a f\'isica n\~ao pode depender delas, mas deve ser universal! Em particular, a f\'isica deve ser descrita e comunicada sem fazer refer\^encia a conven\c c\~oes locais. Essa observa\c c\~ao nos motiva, portanto, a procurar sistemas de unidades de medida que tamb\'em sejam universais e ultrapassem os confins das nossas experi\^encias locais. 

Uma estrat\'egia para achar uma nova base de dimens\~oes que seja universal \'e tentar utilizar leis universais da f\'isica, em particular usando as chamadas {\it constantes fundamentais} que aparecem nessas leis.


\subsection{As tr\^es constantes fundamentais da Natureza} 

H\'a tr\^es constantes fundamentais\footnote{O leitor \'e convidado a ler um divertido debate sobre esse ponto entre tr\^es f\'isicos famosos, cada um com um ponto de vista diferente sobre o assunto \cite{Duff:2001ba}. Nessa revis\~ao, apresentamos, por raz\~oes pedag\'ogicas, o ponto de vista mais comum, das tr\^es constantes fundamentais, que, naquela refer\^encia, \'e defendido pelo L. B. Okun. Foi tamb\'em proposto em \cite{Matsas:2007zz} que seria poss\'ivel medir massas em termos de comprimentos e tempos. Isso rebaixaria a constante de Newton a ser um simples fator de convers\~ao, deixando somente duas grandezas fundamentais, $c$ e $\hbar$.} na Natureza: a {\bf constante de Newton (G)}, a {\bf velocidade da luz (c)} e a {\bf constante de Planck ($\hbar$)}.

A constante de Newton entra ubiquamente nas leis da gravita\c c\~ao, descrevendo o acoplamento do campo gravitacional com a mat\'eria. Por exemplo, entra na lei de gravita\c c\~ao de Newton
\be
F_g=G\frac{m_1 \, m_2}{r^2}\,,
\ee
e tamb\'em nas equa\c c\~oes de Einstein
\be
R_{\mu\nu}-\frac{1}{2}R\, g_{\mu\nu}=\frac{8\pi G}{c^4}T_{\mu\nu}\,.
\ee
Uma das principais caracter\'isticas da gravidade \'e o fato de ser universal, ou seja, de afetar todas as coisas da mesma maneira. Isso se deve, justamente, ao fato de $G$ ser constante e igual para todas as coisas (campos, mat\'eria, part\'iculas, etc.).

A velocidade da luz \'e a velocidade com a qual a luz viaja no v\'acuo e representa a velocidade limite para qualquer objeto no universo, al\'em de ser o fator de convers\~ao entre enegia e massa, conforme a famosa equa\c c\~ao de Einstein $E=m c^2$. 

A constante de Planck representa a unidade m\'inima (ou  {\it quantum}) de momento angular na mec\^anica qu\^antica e entra em muitas equa\c c\~oes, como na radia\c c\~ao de corpo negro ou nos n\'iveis energ\'eticos de \'atomos. Por exemplo, os n\'iveis do \'atomo de hidrog\^enio s\~ao discretos e dados por
\be
E_n=-\frac{m_\mt{e} e^4}{2 \hbar^2 n^2}\,, \qquad n=1,2,\ldots\,.
\ee

O leitor pode encontrar mais detalhes sobre essas tr\^es constantes fundamentais nos livros-texto, como o \cite{moyses}. Nas unidades de medida do Sistema Internacional, essas tr\^es constantes t\^em valores num\'ericos dados por
\bea
G= 6.674 \times 10^{-11}\,\frac{\mbox{m}^3}{\mbox{kg} \cdot \mbox{s}^2}\,,\quad
c= 2.998 \times 10^{8}\,\frac{\mbox{m}}{\mbox{s}}\,,\quad
\hbar= 1.055 \times 10^{-34}\,\frac{\mbox{m}^2\cdot \mbox{kg}}{\mbox{s}}\,.
\eea 
Daqui a pouco, precisaremos desses valores expl\'icitos.


\subsubsection*{Exemplo 6: Raio de Bohr}

Tendo mencionado a constante de Planck, podemos agora considerar uma aplica\c c\~ao muito importante da an\'alise dimensional em mec\^anica qu\^antica: queremos estimar qual \'e o tamanho t\'ipico de um \'atomo (de hidrog\^enio, por simplicidade), mas sem resolver a equa\c c\~ao de Schr\"odinger para o sistema. A primeira coisa a ser feita \'e claramente identificar as grandezas f\'isicas relevantes. Um \'atomo de hidrog\^enio \'e formado por um el\'etron, de massa $m_\mt{e}$ e carga el\'etrica $-e$, e por um pr\'oton, de massa $m_\mt{p}$ e carga $+e$. Parece, ent\~ao, que temos que considerar as seguintes grandezas: $\hbar$ (claro, sendo isso um problema de mec\^anica qu\^antica), o tamanho do \'atomo que queremos achar e que chamamos $a_0$, a carga $e$ e as duas massas $m_\mt{e}$ e $m_\mt{p}$. Uma observa\c c\~ao crucial \'e que, na verdade, a massa que entra nessa conta deveria ser outra, a chamada {\it massa reduzida}\footnote{Quando um problema depende somente da dist\^ancia relativa entre dois objetos, \'e sempre \'util trocar aqueles dois objetos (de massa $m_1$ e $m_2$ e coordenadas $\vec x_1$ e $\vec x_2$) por dois ``objetos virtuais'', um com massa reduzida $\mu = m_1\, m_2/(m_1+m_2)$ e coordenada $\vec r =\vec x_1-\vec x_2$, chamada de {\it relativa}, e um outro de massa igual \`a soma $M=m_1+m_2$ das massas e coordenada $\vec R=(m_1 \vec r_1+m_2 \vec r_2)/M $, chamada de {\it centro de massa}. Toda a din\^amica do problema est\'a contida no objeto com coordenada relativa, enquanto o objeto no centro de massa n\~ao exerce nenhum papel interessante.}
\be
\mu=\frac{m_\mt{e} \, m_\mt{p}}{m_\mt{e}+m_\mt{p}}\,.
\ee 
Sendo o pr\'oton muito mais pesado que o el\'etron, podemos aproximar $\mu\simeq m_\mt{e}$. Portanto, a massa do pr\'oton n\~ao \'e de verdade relevante, pelo menos em uma primeira estimativa.\footnote{Refinando as coisas, podemos pensar que $a_0$ depende tamb\'em de uma fun\c c\~ao $f(m_\mt{e}/m_\mt{p})$ da raz\~ao das massas, que, por\'em, n\~ao pode ser determinada pela an\'alise dimensional, por ser adimensional.} A combina\c c\~ao de $\hbar$, $e$ e $m_\mt{e}$ com dimens\~oes de comprimento resulta ser
\be
a_0=\frac{\hbar^2}{m_\mt{e} e^2}\,,
\label{Bohrradius}
\ee 
como pode ser visto usando o procedimento ilustrado na Se\c c\~ao \ref{proc}. Esse comprimento \'e chamado de {\it raio de Bohr} e, de fato, \'e a quantidade que define o tamanho t\'ipico dos orbitais eletr\^onicos, como pode ser visto explicitamente resolvendo a equa\c c\~ao de Schr\"odinger. Usando os valores expl\'icitos no Sistema Internacional, resulta que $a_0\simeq 5.29\times 10^{-11}$ m.

\begin{figure}
\begin{center}
\setlength{\unitlength}{1cm}
\includegraphics[scale=.6]{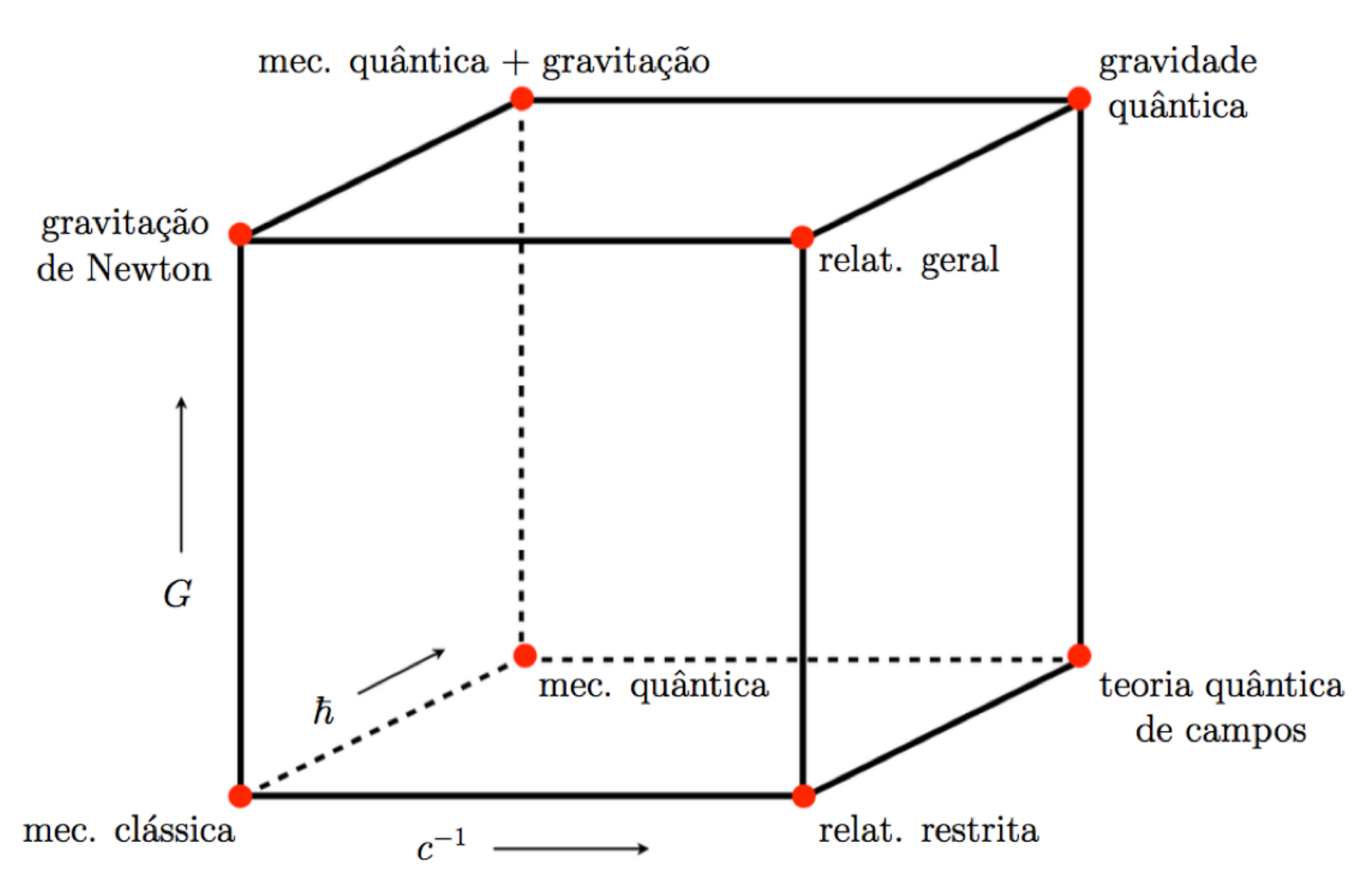} 
\end{center}
\caption{\small
O cubo da f\'isica \cite{Okun:2001rd}.}
\label{cubo}
\end{figure}

\subsection{O cubo da f\'isica}

\'E muito interessante observar que as tr\^es constantes introduzidas acima oferecem um jeito muito legal de organizar toda a f\'isica em um diagrama chamado de {\it cubo da f\'isica}.\footnote{Ou {\it cubo de Bronshtein-Zelmanov-Okun}, ver por exemplo \cite{Okun:2001rd}.} Ver a Figura~\ref{cubo}.
Consideremos, inicialmente, o caso em que n\~ao haja essas constantes, ou seja ``$G\to 0$, $\hbar\to 0$, $c\to \infty$''.\footnote{Como deve ter ficado claro pela discuss\~ao anterior, isso n\~ao \'e estritamente correto, pois $G$, $c$ e $\hbar$ s\~ao dimensionais, enquanto 0 ou $\infty$ s\~ao adimensionais. O limite significa que $G$ e $\hbar$ s\~ao muito menores que qualquer outra grandeza com as mesmas dimens\~oes no problema (e \'e a raz\~ao pela qual coloquei as aspas) e que $c$ \'e muito maior de qualquer outra velocidade no problema.} Esse \'e o caso da mec\^anica de Newton sem a gravidade. ``Ligando'' somente $G$, achamos a gravita\c c\~ao de Newton; ligando somente $c$, achamos a relatividade restrita; ligando somente $\hbar$, achamos a mec\^anica qu\^antica.

Podemos agora ligar dois constantes ao mesmo tempo. Ligando $G$ e $c$, achamos a relatividade geral de Einstein; ligando $G$ e $\hbar$, achamos a mec\^anica qu\^antica n\~ao relativ\'istica mais gravita\c c\~ao; ligando $c$ e $\hbar$, achamos a teoria qu\^antica de campos. O  \'ultimo desafio da f\'isica te\'orica, o Santo Graal, \'e entender o caso quando todas as tr\^es constantes s\~ao ligadas, ou seja a {\it gravidade qu\^antica}, que \'e o casamento entre mec\^anica qu\^antica e relatividade geral.


\subsection{O cerne da teoria qu\^antica de campos}   

Motivados pela apari\c c\~ao da teoria qu\^antica de campos no cubo da f\'isica, vamos gastar duas palavras para esbo\c car a ideia principal dessa teoria. A uni\~ao entre a relatividade restrita e a mec\^anica qu\^antica tem uma consequ\^encia crucial: part\'iculas podem nascer e morrer. Em outras palavras, o n\'umero de part\'iculas no problema n\~ao \'e constante, ou {\it conservado}, como acontece na mec\^anica cl\'assica ou na mec\^anica qu\^antica sem relatividade restrita. 

\begin{figure}
\begin{center}
\begin{tabular}{cc}
\setlength{\unitlength}{1cm}
\hspace{-0.9cm}
\includegraphics[width=4.5cm]{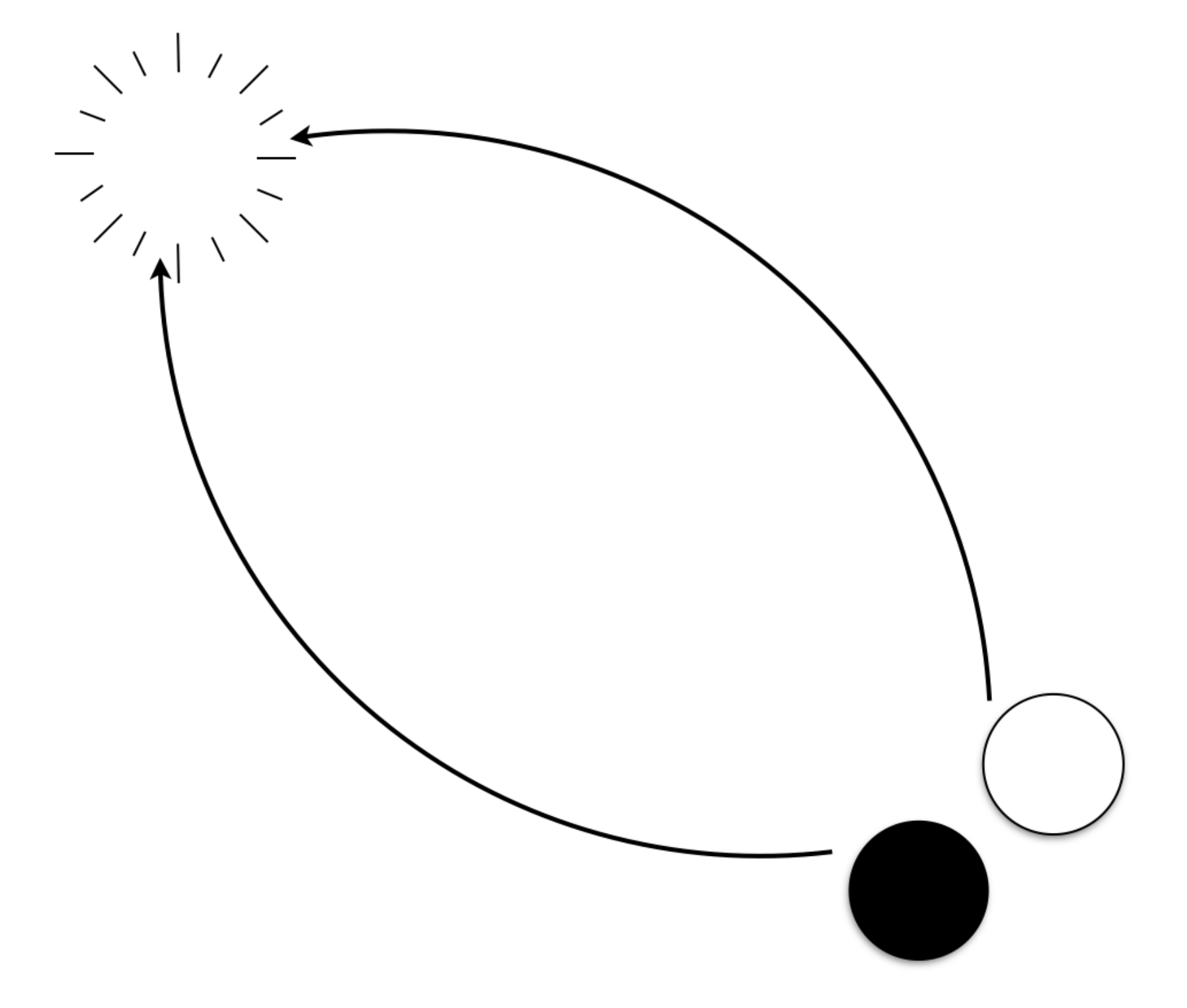} 
\qquad\qquad & 
\includegraphics[width=7.5cm]{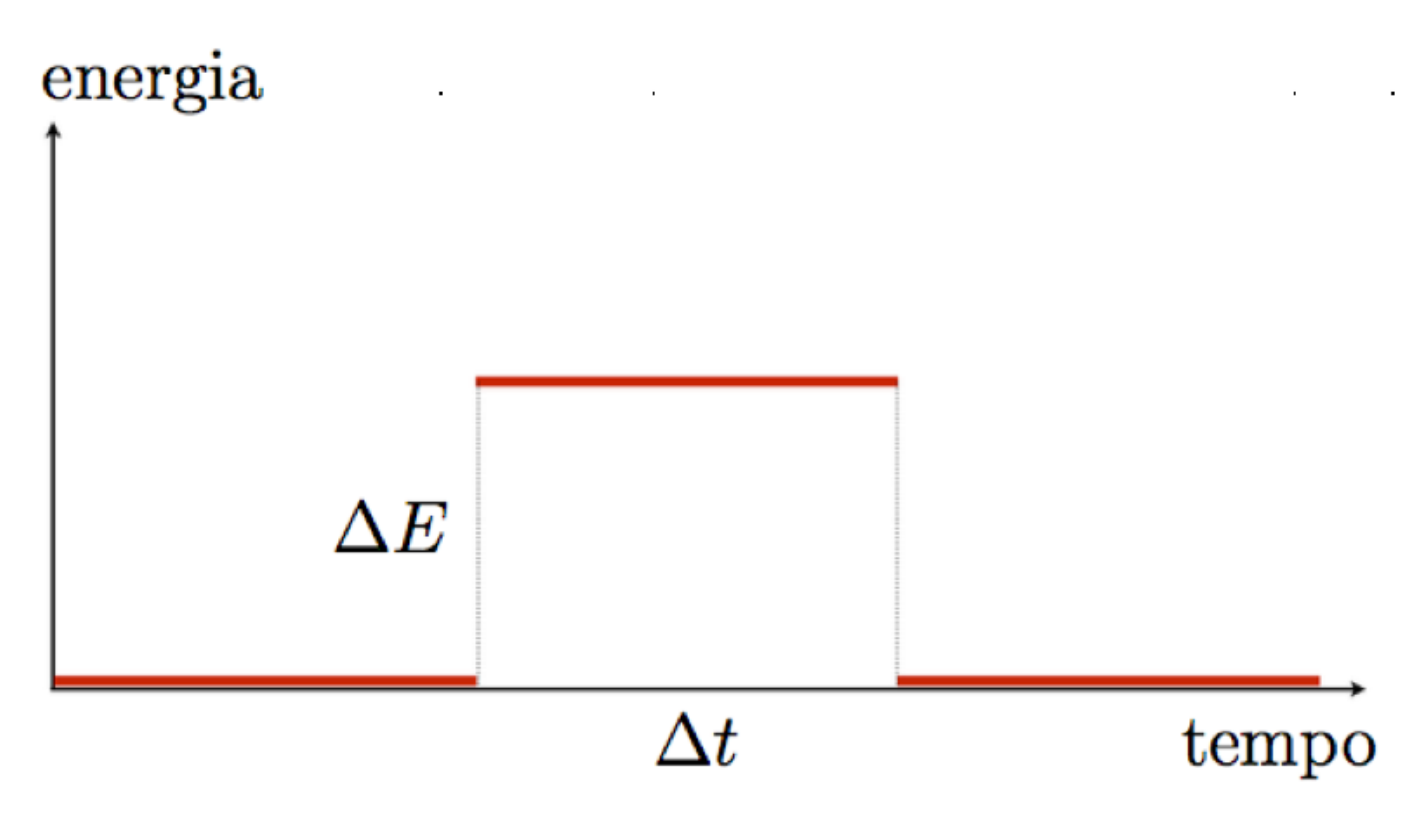}
\qquad
\end{tabular}
\end{center}
\caption{\small
(Esq.) Produ\c c\~ao e aniquila\c c\~ao no v\'acuo de um par part\'icula/anti-part\'icula. (Dir.) A conserva\c c\~ao de energia pode ser ``violada'' durante breves intervalos de tempo, gra\c cas ao princ\'ipio de indetermina\c c\~ao entre energia e tempo.}
\label{TQC}
\end{figure}

Na relatividade restrita, a energia \'e equivalente \`a massa, conforme a f\'ormula mencionada acima: $E=m c^2$. Tendo energia, podemos criar massa, ou seja, podemos criar part\'iculas. Na teoria qu\^antica de campos, o {\it v\'acuo}, que \'e o estado de menor energia, n\~ao \'e vazio, mas repleto de part\'iculas e anti-part\'iculas {\it virtuais}. Podemos excitar essas part\'iculas e anti-part\'iculas virtuais e convert\^e-las em part\'iculas e anti-part\'iculas {\it reais}, tendo energia a disposi\c c\~ao, por exemplo, usando aceleradores como o LHC. Tamb\'em \'e poss\'ivel, por\'em, que pares de part\'iculas e anti-part\'iculas\footnote{Devem ser pares de part\'iculas e correspondentes anti-part\'iculas (part\'iculas de carga oposta) para respeitar crit\'erios de conserva\c c\~ao, como a conserva\c c\~ao da carga el\'etrica, que no v\'acuo \'e zero.} surjam do v\'acuo espontaneamente para, logo depois, se aniquilarem, ver a Figura~\ref{TQC} (Esq.).
Claramente, isso seria uma viola\c c\~ao da conserva\c c\~ao da energia, pois o v\'acuo tem energia zero, enquanto o par tem energia igual, pelo menos, \` a massa deste, que \'e duas vezes a massa da part\'icula, pois part\'iculas e correspondentes anti-part\'iculas t\^em a mesma massa. Ver a Figura~\ref{TQC} (Dir.). A resolu\c c\~ao desse paradoxo est\'a na mec\^anica qu\^antica, em particular, na rela\c c\~ao de indetermina\c c\~ao de Heisenberg
\be
\Delta E \, \Delta t\geq \frac{\hbar}{2}\,,
\ee
que n\~ao nos permite medir a energia com precis\~ao arbitr\'aria em intervalos finitos de tempo. Podemos tolerar uma viola\c c\~ao da conserva\c c\~ao da energia igual a $\Delta E$ para um (curto) intervalo de tempo $\Delta t\leq \frac{\hbar}{2\Delta E}$, que \'e justamente o intervalo de tempo que a part\'icula e a anti-part\'icula vivem antes de se aniquilarem uma com a outra. 


\subsubsection*{Exemplo 7: Constante de estrutura fina}

Consideremos, agora, el\'etrons em teoria qu\^antica de campos. As grandezas dimensionais \`a nossa disposi\c c\~ao s\~ao: a constante de Planck $\hbar$, a velocidade da luz $c$, a carga do el\'etron $e$ e a massa do el\'etron $m_\mt{e}$. A pergunta \'obvia \'e a seguinte: podemos construir uma combina\c c\~ao adimensional $Y$ dessas grandezas? Para fazer isso, escrevemos
\be
Y = \hbar^{a_1} c^{a_2} e^{a_3} m_\mt{e}^{a_4} \quad \to 
\quad [Y]=1= M^{a_1+\frac{1}{2}a_3+a_4} L^{2a_1+a_2+\frac{3}{2}a_3} T^{-a_1-a_2-a_3}\,,
\ee 
onde usamos que a carga el\'etrica tem dimens\~oes de $\sqrt{ML^3T^{-2}}$. Uma solu\c c\~ao \'e $(a_1=-1, a_2=-1, a_3=2,a_4=0)$ e a correspondente combina\c c\~ao adimensional \'e dada por
\be
\frac{e^2}{\hbar \, c}=7.297 \times 10^{-3}\simeq \frac{1}{137}\,.
\ee
A observa\c c\~ao importante \'e que esse \'e um n\'umero pequeno. O leitor deve lembrar agora a terceira observa\c c\~ao na Se\c c\~ao~\ref{proc}: n\'umeros pequenos e adimensionais podem ser usados como par\^ametros nas expans\~oes. De fato, o n\'umero acima \'e chamado {\it constante de estrutura fina}\footnote{Na verdade, n\~ao \'e constante e depende da energia do sistema, mas isso \'e uma outra hist\'oria!} e aparece nas expans\~oes {\it perturbativas} de algumas teorias de campos, como a eletrodin\^amica qu\^antica (QED).


\section{Sistema de unidades naturais}
\label{planck}

O leitor atento deve ter observado que as tr\^es constantes introduzidas acima t\^em dimens\~oes independentes:
\be
[G]=M^{-1}L^3 T^{-2}\,,\qquad 
[c]=L T^{-1}\,,\qquad 
[\hbar]=M L^2 T^{-1}\,.
\ee
\'E poss\'ivel, portanto, us\'a-las para definir uma nova base de dimens\~oes, as dimens\~oes de $[G]$, $[c]$ e $[\hbar]$. Daqui, podemos mudar de base de novo e voltar a uma base de $M$, $L$ e $T$. Explicitamente, vamos escrever
\be
[G]^{a_1} [c]^{a_2} [\hbar]^{a_3}=M^{-a_1+a_3}L^{3a_1+a_2+2a_3}
T^{-2a_1-a_2-a_3}\,.
\ee 
Agora, queremos achar combina\c c\~oes com dimens\~oes de massa, comprimento e tempo:
\bea
\begin{array}{ccc}
\mbox{massa}  & \hskip2cm  \mbox{comprimento} \hskip 2cm &\mbox{tempo}\\
-a_1+a_3=1 &  -a_1+a_3=0 & -a_1+a_3=0 \\
3a_1+a_2+2a_3=0 &3a_1+a_2+2a_3=1 & 3a_1+a_2+2a_3=0\\
-2a_1-a_2-a_3=0 & -2a_1-a_2-a_3=0 & -2a_1-a_2-a_3=1\,.
\end{array}
\eea
As solu\c c\~oes s\~ao chamadas {\it massa de Planck} $m_\mt{P}$, {\it comprimento de Planck} $\ell_\mt{P}$ e {\it tempo de Planck} $t_\mt{P}$:
\bea
m_\mt{P} = \sqrt{\frac{\hbar \, c}{G}}\,,\qquad \ell_\mt{P}=\sqrt{\frac{G\, \hbar}{c^3}}\,,\qquad t_\mt{P}=\sqrt{\frac{G\, \hbar}{c^5}}\,.
\eea
O ponto crucial desse exerc\'icio \'e que a massa, o comprimento e o tempo de Planck t\^em um car\'ater universal, pois foram definidas a partir de constantes fundamentais! Eles representam as escalas de massa (de objetos elementares), comprimento e tempo onde efeitos de gravidade qu\^antica viram importantes. Em outras palavras, quando consideremos fen\^omenos que acontecem em comprimentos de Planck, ou em escalas de tempo de Planck, o nosso entendimento usual de espa\c co-tempo da relatividade geral deve ser modificado, para incluir efeitos qu\^anticos. Isso tem a ver com o \'ultimo v\'ertice do cubo da f\'isica, na Figura~\ref{cubo}.

Os valores dessas quantidades no Sistema Internacional s\~ao dados por
\be
m_\mt{P}=2.176 \times 10^{-5}\mbox{ g}\,,\qquad \ell_\mt{P}=1.616 \times 10^{-33}\mbox{ cm}\,,\qquad
t_\mt{P}=5.391 \times 10^{-44}\mbox{ s}\,.
\ee
Reparamos que esses valores s\~ao extremos! O comprimento de Planck e o tempo de Planck s\~ao extremamente menores que todos os comprimentos e tempos aos quais estamos acostumados. A massa de Planck \'e, por outro lado, muito maior que a massa das part\'iculas elementares (por exemplo, \'e 19 ordens de magnitude maior que a massa do pr\'oton).\footnote{O ponto aqui \'e que devemos comparar a massa de Planck, que \'e algo fundamental, com a massa de coisas tamb\'em fundamentais, ou elementares, e n\~ao compostas. Coisas compostas podem claramente ter massas bem maiores que $m_\mt{P}$!}

Agora, \'e natural ir um passo adiante e definir um novo sistema de unidades, chamado de {\it unidades naturais} ou {\it unidades de Planck}. Nesse sistema declaramos que
\be
c\to1\,,\qquad  \hbar\to 1\,,\qquad G\to 1\,. 
\ee
Vamos entender o que isso significa. Declarar que a velocidade da luz tende a 1, ou seja, um par\^ametro adimensional, implica em comprimentos e tempos com a mesma dimens\~ao. Isso deve ser familiar para o leitor: um {\it ano-luz} \'e uma medida de comprimento, embora a palavra ``ano'' seja usada na express\~ao, e \'e a dist\^ancia que a luz viaja em um ano ($\simeq 9.46\times 10^{12}$ km). Velocidades s\~ao, portanto, medidas em rela\c c\~ao \`a velocidade da luz e variam de 0 a 1, que \'e a velocidade limite. Outra consequ\^encia \'e que $E=m c^2$ vira $E=m$, ou seja, energia e massa v\~ao tamb\'em ter as mesmas dimens\~oes.

Quando $c\to 1$, a constante de Planck $\hbar$ vai ter dimens\~oes de $M L$. Declarar que a mesma $\hbar$ vira 1 significa portanto que massas (e energias) viram inversos de comprimentos. Por fim, declarar que $G$ vira 1 significa que a massa ou o comprimento de refer\^encia s\~ao 1.

Notamos que, na pr\'atica, fala-se que $c=1$, $\hbar=1$ e $G=1$, em vez de colocar o s\'imbolo de limite, mas o significado \'e o mesmo. Vale a pena ressaltar um ponto que, \`as vezes, confunde os alunos: colocar $c=1$, $\hbar=1$ e $G=1$ n\~ao significa ``perder informa\c c\~oes'' sobre as grandezas e dimens\~oes do problema! Querendo, podemos sempre inserir univocamente de volta $c$, $\hbar$ e $G$ nas f\'ormulas, somente usando a an\'alise dimensional.
 

\subsubsection*{Exemplo 8: Radia\c c\~ao Hawking e entropia de buracos negros}

Buracos negros est\~ao entre os objetos mais interessantes e misteriosos do universo. Os buracos negros mais simples, presentes na relatividade geral, s\~ao chamados de Schwarzschild.\footnote{Os buracos negros astrof\'isicos n\~ao s\~ao de Schwarzschild, pois geralmente t\^em momento angular e rodam.} S\~ao est\'aticos e esf\'ericos com um raio chamado {\it raio de Schwarzschild} e dado por 
\be
R_\mt{S}= \frac{2 GM}{c^2}\,,
\label{Sch}
\ee
onde $M$ \'e a massa do buraco negro. Para ter uma ideia, um buraco negro com massa igual \`a massa do Sol tem raio $R_\mt{S}\simeq 3$ km. O leitor deve ter ouvido falar que buracos negros s\~ao t\~ao densos que deformam o espa\c co-tempo ao seu redor, ao ponto de n\~ao deixar escapar nada, nem a luz, que entra no raio de Schwarzschild (onde h\'a o chamado {\it horizonte de eventos}). De fato, isso \'e verdade somente se desprezarmos efeitos qu\^anticos. Bekenstein e Hawking demonstraram, nos anos '70, que buracos negros emitem uma radia\c c\~ao, e portanto t\^em propriedades termodin\^amicas, como temperatura e entropia. Para mais detalhes, ver, por exemplo, o \'otimo livro-texto \cite{Zee:2013dea}. 

\begin{figure}
\begin{center}
\setlength{\unitlength}{1cm}
\includegraphics[scale=.5]{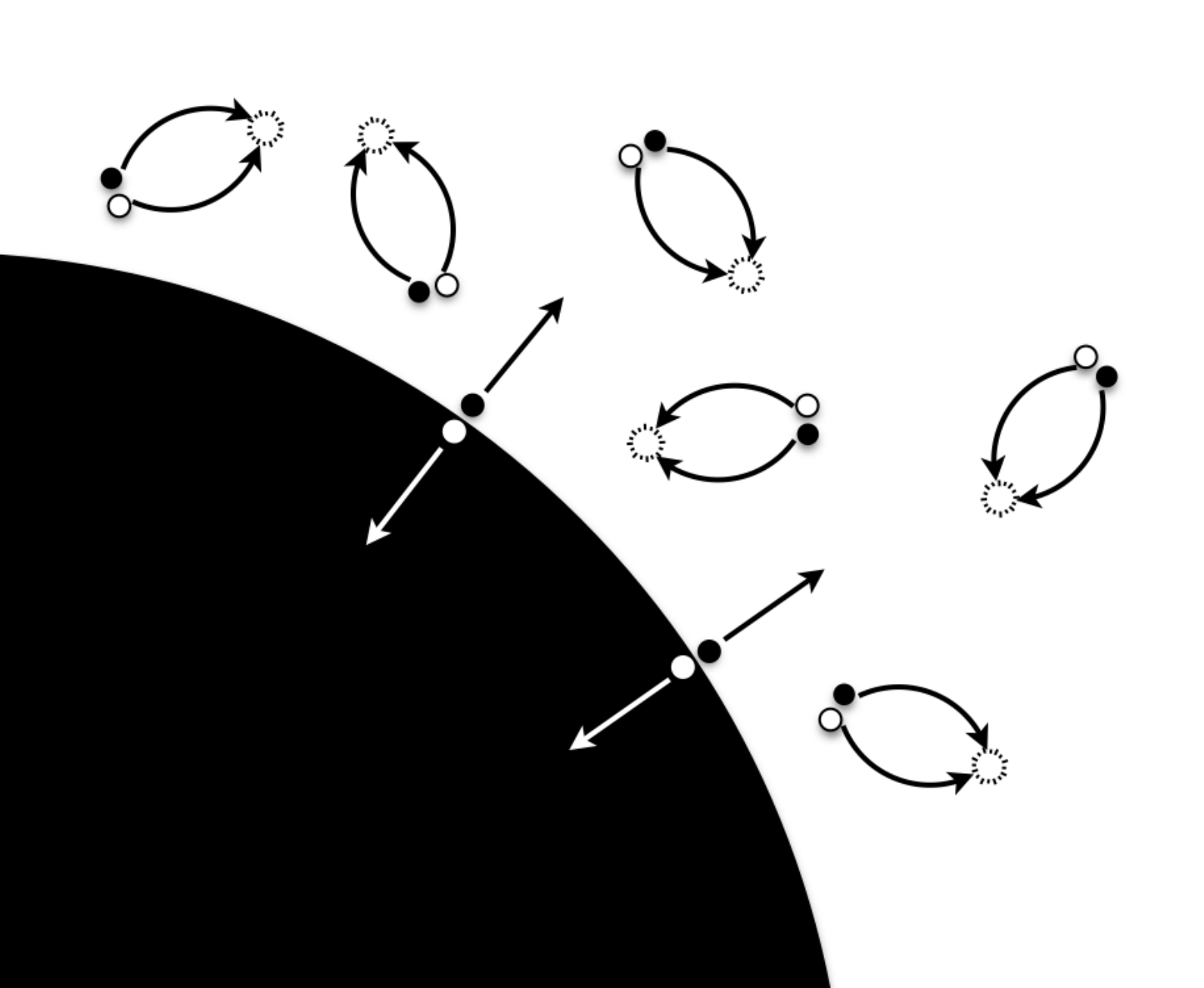} 
\end{center}
\caption{\small
A radia\c c\~ao Hawking de um buraco negro. As part\'iculas virtuais que nascem perto do horizonte de eventos s\~ao separadas das pr\'oprias parceiras e viram reais.}
\label{BH}
\end{figure}

Uma maneira heur\'istica de entender a fonte dessa radia\c c\~ao (chamada de {\it radia\c c\~ao Hawking}) \'e lembrar a hist\'oria das part\'iculas da teoria de campos que surgem do v\'acuo e se aniquilam. Se o horizonte de eventos se encontra exatamente entre a part\'icula e a anti-part\'icula que surgiram do v\'acuo, a part\'icula dentro do horizonte ser\'a absorvida pelo buraco negro, enquanto a part\'icula fora do horizonte deixa de ser virtual para virar real e se propagar como uma part\'icula ordin\'aria (pois ela n\~ao tem mais um parceiro para se aniquilar). Ver a Figura~\ref{BH}. A radia\c c\~ao Hawking pode ent\~ao ser pensada como se fosse constitu\'ida por essas part\'iculas que, de virtuais, viraram reais. 

\'E poss\'ivel estimar a temperatura e entropia do buraco negro usando a an\'alise dimensional \cite{Zee:2013dea}. Adotamos um sistema de unidades com $c =1=\hbar$, mas $G\neq 1$. Os par\^ametros do problema s\~ao, portanto, $G$ e $M$ (e $R_\mt{S}$ que, por\'em, depende desses dois). Nessas unidades, a combina\c c\~ao $GM$ tem dimens\~oes de comprimento ou de inverso de massa. Lembrando que temperatura \'e energia (a energia m\'edia dos graus de libertade microsc\'opicos), a temperatura do buraco negro deve ser dada por
\be
T\sim \frac{1}{GM}\,.
\label{tempH}
\ee
Fazendo a conta detalhada e colocando todos os fatores, dimensionais e adimensionais, seria $T=\hbar c^3/8\pi GM$. Vemos, portanto, que: 1) essa radia\c c\~ao \'e, de fato, um efeito qu\^antico, pois se anula quando $\hbar$ \'e desprez\'ivel; 2) \'e explosiva para massas pequenas! \'E interessante frisar que, para um buraco negro com massa solar, a temperatura \'e muito baixa: $T\simeq 10^{-7}$ K.

Conforme a primeira lei da termodin\^amica, $dE=T\, dS$, associada a uma temperatura h\'a uma entropia. Lembrando que a massa do buraco negro \'e a energia (pois $c=1$), podemos integrar essa lei usando (\ref{tempH})
\be
\frac{dE}{dS}=\frac{dM}{dS}\sim \frac{1}{GM}\quad \to
\quad S\sim GM^2\,.
\ee
Para uma massa solar, isso d\'a uma entropia enorme: $S\simeq 10^{77}$.  Em termos do raio de Schwarzschild, podemos re-expressar a entropia como
\be
S\sim \frac{R^2_\mt{S}}{G}\,,
\ee
ou seja, a entropia $S$  de um buraco negro \'e proporcional \`a {\bf \'area do horizonte}, que \'e uma esfera, $A=4\pi R^2_\mt{S}$.\footnote{Colocando todos os fatores, ter\'iamos $S=k_\mt{B} c^3 A/4 G$, com $k_\mt{B}$ a constante de Boltzmann.} Isso \'e muito diferente do que acontece em sistemas termodin\^amicos usuais, onde a entropia \'e extensiva e varia com o {\it volume} da \'area ocupada, e n\~ao com a \'area da borda desse volume! Isso tem consequ\^encias muito profundas. As mais importantes s\~ao, talvez, o {\it princ\'ipio hologr\'afico} \cite{suss} e a {\it correspond\^encia AdS/CFT}~\cite{Maldacena}. 

Lembrando a discuss\~ao acima sobre a cria\c c\~ao de part\'iculas em aceleradores, podemos, agora, dar uma interpreta\c c\~ao muito interessante ao comprimento de Planck como {\it comprimento m\'inimo}. A ideia \'e que, em um mundo sem gravidade, podemos, em princ\'ipio, sondar dist\^ancias arbitrariamente pequenas aumentando a energia do acelerador. Conforme a mec\^anica qu\^antica, com energia $E$ do feixe podemos sondar part\'iculas localizadas em uma regi\~ao chamada {\it comprimento de de Broglie}, dada por 
\be
\ell_\mt{dB}\sim \frac{1}{E}\,,
\ee
em unidades de $c=\hbar=1$. Introduzindo a gravidade, por\'em, o cen\'ario muda drasticamente, pois h\'a buracos negros. Uma energia $E$ concentrada dentro de um raio de Schwarzschild (\ref{Sch}) com $M\sim E$ vai, de fato, colapsar em um buraco negro. O feixe do acelerador vai formar um buraco negro quando 
\be
R_\mt{S}\sim GE \gtrsim \ell_\mt{dB} \sim \frac{1}{E} \quad \to \quad E\gtrsim \sqrt{\frac{1}{G}}=m_\mt{P}\,,
\ee 
sempre em unidades de $c=1=\hbar$. Portanto, dist\^ancias menores que $\ell_\mt{P}$ n\~ao s\~ao acess\'iveis, pois um buraco negro de $R_\mt{S}\gtrsim \ell_\mt{P}$ \'e formado no processo e n\~ao podemos olhar dentro de um buraco negro!

Vamos terminar essa se\c c\~ao com um simples exerc\'icio \cite{zwiebach}. A ideia \'e se convencer que efeitos de gravidade qu\^antica s\~ao desprez\'iveis em situa\c c\~oes ordin\'arias. J\'a vimos, por exemplo, que a temperatura da radia\c c\~ao Hawking \'e extremamente baixa para buracos negros com massas compar\'aveis \`a do Sol. Alguns buracos negros astrof\'isicos chegam a ter mais que $10^{10}$ massas solares. Para eles, essa temperatura seria ainda menor, pois varia com o inverso da massa. 

Agora queremos estimar a import\^ancia da intera\c c\~ao gravitacional em \'atomos. Qual seria o raio de Bohr do \'atomo de hidrog\^enio se a intera\c c\~ao entre el\'etron e pr\'oton fosse somente gravitacional e n\~ao eletromagn\'etica? Para resolver isso, lembramos que, na lei de Coulomb (em unidades e.s.u., sem $1/4\pi\epsilon_0$), a intensidade da intera\c c\~ao varia com $-e^2$, ou seja, o produto das cargas de el\'etron e pr\'oton. Por outro lado, na lei de gravita\c c\~ao de Newton, o acoplamento gravitacional varia com $G$ vezes o produto das massas: $G m_\mt{e} m_\mt{p}$. Portanto, para achar o ``raio de Bohr gravitacional'' devemos trocar esses acoplamentos em (\ref{Bohrradius}):
\be
a_0=\frac{\hbar^2}{m_\mt{e} e^2} \quad \to \quad a_0^{(g)}=\frac{\hbar^2}{m_\mt{e}(G m_\mt{e} m_\mt{p})}
\simeq 1.20 \times 10^{31} \mbox{ cm}\,.
\ee
A intera\c c\~ao seria t\~ao pequena que o raio do \'atomo resultaria ser maior que o universo observ\'avel! Isso \'e uma consequ\^encia direta do fato de $G$ ser t\~ao pequena.

Efeitos de gravidade qu\^antica devem ser inclu\'idos em situa\c c\~oes extremas, como perto das singularidades no centro de buracos negros ou no estudo do universo primordial, quando temperatura e densidade eram alt\'issimas, da ordem da escala de Planck.  


\section{Conclus\~ao}    

Espero ter convencido o leitor sobre o poder e a eleg\^ancia da an\'alise dimensional: em muitas situa\c c\~oes, podemos ter indica\c c\~oes sobre quais s\~ao as escalas e as combina\c c\~oes de par\^ametros relevantes em um problema, simplesmente analisando as grandezas dimensionais do mesmo. Vimos isso em v\'arios exemplos de v\'arias \'areas da f\'isica, da mec\^anica b\'asica at\'e a efeitos qu\^anticos na gravidade. 

Agora, \`as vezes, a an\'alise dimensional e as expectativas baseadas em compara\c c\~oes com escalas naturais em um problema fracassam tragicamente: \'e o caso dos chamados {\it problemas de hierarquias}. H\'a muitas hierarquias n\~ao ainda explicadas e que representam algumas das perguntas em aberto mais importantes da f\'isica. Por exemplo, vimos que a massa do pr\'oton \'e muito menor da massa de Planck. Por qu\^e? O exemplo arquet\'ipico desses problemas \'e o problema da {\it constante cosmol\'ogica} respons\'avel pela expans\~ao acelerada do universo: a discrep\^ancia entre expectativa e medi\c c\~ao \'e de 120 ordens de magnitude naquele caso! 

A resposta a qualquer uma dessas perguntas seria, com certeza, digna de um pr\^emio Nobel. Vamos deixar uma discuss\~ao mais aprofundada sobre esses temas para uma outra ocasi\~ao.


\subsection*{Agradecimentos}
Agrade\c co a Prof.a Renata Zukanovich Funchal e toda a organiza\c c\~ao do {\it Convite \`a F\'isica} do Instituto de F\'isica da USP, pelo convite a apresentar esse material em uma palestra em agosto de 2015. Agrade\c co tamb\'em a Deborah Liguori para a ajuda com a revis\~ao do texto. A minha pesquisa \'e financiada em parte pelo CNPq e em parte pela FAPESP (projetos 2014/18634-9 e 2015/17885-0). Coment\'arios e pedidos de esclarecimentos s\~ao muito bem vindos.


\end{document}